\documentclass[floatfix,twocolumn,aps,prd,showpacs,amsmath,amssymb]{revtex4-1}
\usepackage[latin1]{inputenc}
\usepackage[dvips]{graphicx}

\usepackage{dcolumn}
\usepackage{bm}
\usepackage{dsfont}
\usepackage{color}
\usepackage{ulem} 
\usepackage{url}
\usepackage[colorlinks=true ,citecolor=blue]{hyperref}
\usepackage{amsmath,amssymb}

\newcommand{\be}{\begin{equation}}
	\newcommand{\ee}{\end{equation}}
\newcommand{\bq}{\begin{eqnarray}}
	\newcommand{\eq}{\end{eqnarray}}

\newcommand{\PR}[1]{\ensuremath{\left[#1\right]}}

\begin{document}

\title{Chern-Simons theory and atypical Hall conductivity in the Varma phase}

\author{Nat\'{a}lia Menezes$^{1}$, Cristiane Morais Smith$^{1}$ and Giandomenico Palumbo$^{1}$}
\affiliation{$^1$Institute for Theoretical Physics, Center for Extreme Matter and Emergent Phenomena, Utrecht University, Princetonplein 5, 3584CC Utrecht, the Netherlands}

\date{\today}

\begin{abstract}
In this letter, we analyze the topological response of a fermionic model defined on the Lieb lattice in presence of an electromagnetic field. The tight-binding model is built in terms of three species of spinless fermions and supports a topological Varma phase due to the spontaneous breaking of time-reversal symmetry.
In the low-energy regime, the emergent effective Hamiltonian coincides with the so-called Duffin-Kemmer-Petiau (DKP) Hamiltonian, which describes relativistic pseudospin-0 quasiparticles. By considering a minimal coupling between the DKP quasiparticles and an external Abelian gauge field, we first find the Landau-level spectrum by fixing the Landau gauge; then we compute the emergent Chern-Simons theory for a weak-electromagnetic-field regime. The corresponding Hall conductivity reveals an atypical quantum Hall effect, which can be simulated in an artificial Lieb lattice.
 \end{abstract}

\maketitle

{\bf Introduction:--} In 1928, Dirac proposed a theory for the description of relativistic spin-1/2 particles. Since then, his model has found several theoretical applications, which have extrapolated the boundaries between high- and low-energy physics. Nowadays, the Dirac theory plays a fundamental role in condensed-matter physics on the description of materials such as graphene \cite{Novoselov}, topological insulators \cite{Kane}, and topological superconductors \cite{Zhang}, due to their relativistic-like bulk/edge band dispersion. These so-called Dirac materials have unique properties, like the bulk-edge correspondence \cite{Fradkin}, which allow for innovative applications in future quantum technologies \cite{MacDonald,Pachos}. Differently from the fundamental fermions in the original model \cite{Ramond}, the effective low-energy theory in these materials describes either Dirac or Weyl quasiparticles that emerge in their bulk and/or edge. For instance, in three dimensions, massless Dirac and Weyl quasiparticles appear in the bulk of Dirac \cite{Rappe} and Weyl semimetals \cite{Turner}, respectively. Moreover, there exists a deep connection between the Dirac theory and topology via the Clifford bundles, K-theory and the index theorem \cite{Nakahara}, which has allowed to classify, in a mathematical way, topological insulators, topological superconductors and topological semimetals in any dimension \cite{Altland, Ryu2, Kitaev, Ryu}.

Dirac materials, however, represent only a particular class of quantum systems with relativistic band dispersion. Several models described by \textit{quasi-relativistic} non-Dirac Hamiltonians have been recently proposed both in condensed-matter and cold-atom systems \cite{Bercioux,Shen,Green,Franz,Moessner,Goldman,Goldman2,Morais-Smith,Orlita,Bernevig,Peotta,Soluyanov,Assaad,Stern}. In these works, quasiparticles carrying pseudospin-1 \cite{pseudo1} emerge in the bulk of the material and are associated to the presence of flat bands on the Lieb lattice.
This lattice has been recently implemented in artificial electronic \cite{Morais-Smith2,Ojanen} and photonic \cite{Vicencio,Goldman3,Diebel} systems, opening the way to experimentally investigating novel quantum phenomena related to the presence of quasiparticles with integer pseudospin.

Furthermore, unlike the previously mentioned works, a novel topological system on the Lieb lattice supporting \textit{relativistic} pseudospin-0 quasiparticles has been theoretically proposed in Ref.~\cite{Palumbo}. These quasiparticles are described by a two-dimensional Duffin-Kemmer-Petiau (DKP) theory \cite{Nieto}, which represents an extension of the Dirac theory for integer-spin particles. This model gives rise to a Chern-semimetallic phase, where the chiral edge modes are topologically protected by a non-zero Chern number in the bulk.

\begin{figure}[htp]
	\begin{center}
		\includegraphics[scale=0.20]{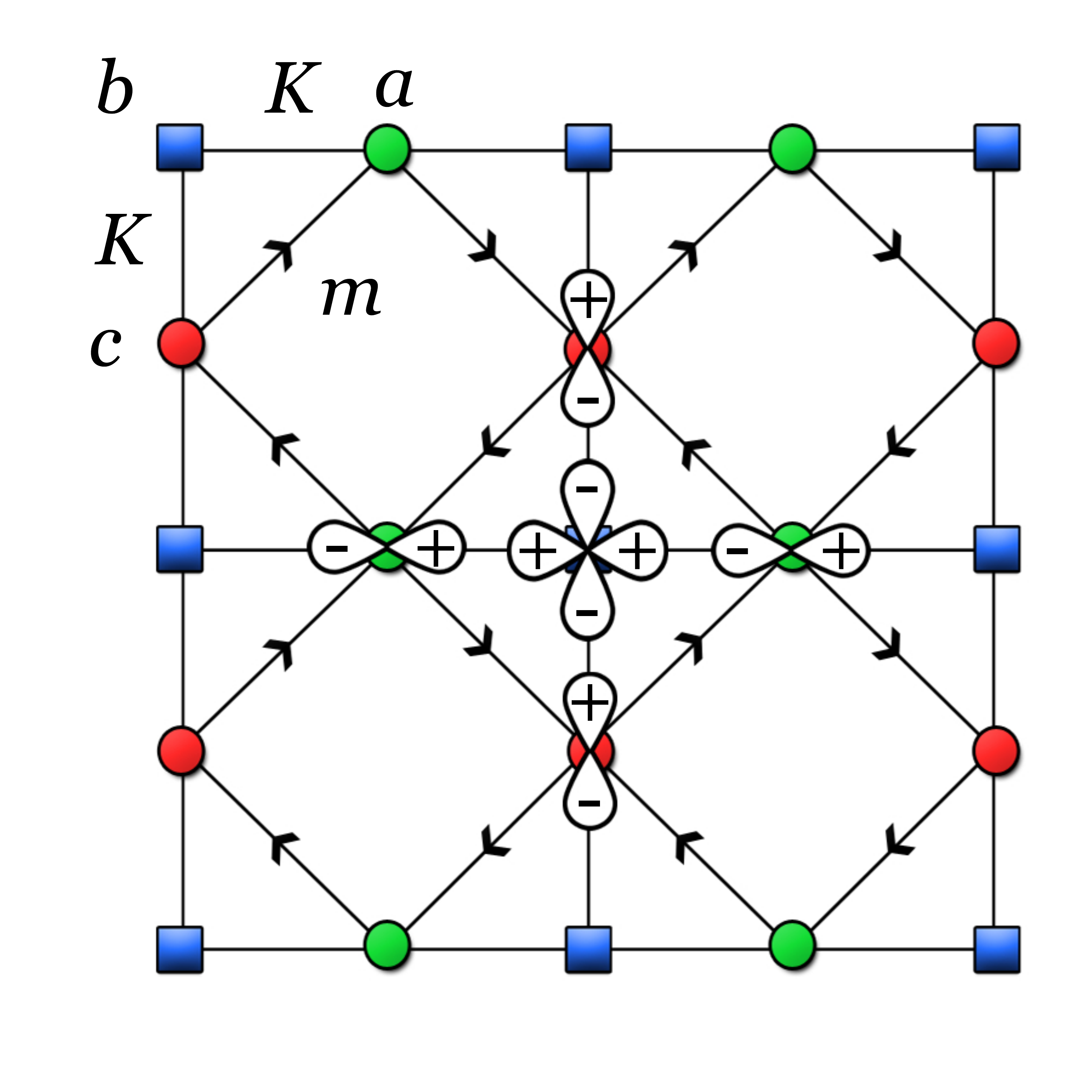}
	\end{center}
	\caption{\label{Fig1} Lieb lattice with three species of spinless fermions labeled by $a$, $b$ and $c$, which are associated to $p_{x}$, $d_{x^{2}-y^{2}}$ and $p_{y}$ orbitals, respectively. The tunneling $K$ is induced by the overlap between $d_{x^{2}-y^{2}}$ and $p_{x}$ ($p_{y}$) orbitals in $x$($y$)-direction.
		The diagonal lines correspond to the next-nearest-neighbor (NNN) loop currents between $p_{x}$ and $p_{y}$ orbitals, that arises from a mean-field description, with $m$ the amplitude of the complex hopping.}
\end{figure}

From a more concrete perspective, the Lieb lattice is relevant in the study of high-Tc cuprates, where the Varma phases \cite{Varma,Varma1} were shown to be important in the description of the pseudogap regime. Within this picture, the copper (Cu) sites are characterized by $d_{x^{2}-y^{2}}$ orbitals and the planar oxygen (O) by $p_{x}$ ($p_{y}$) orbitals. Due to interaction, orbital current loops emerge, spontaneously breaking time-reversal symmetry, but preserving the translational symmetry of the lattice (see Fig.~\ref{Fig1}).

The aim of this work is to study the topological response of the Varma phase introduced in Ref.~\cite{Moore-Varma} to an applied electromagnetic field. We first show that the low-energy theory coincides with the two-dimensional DKP theory by unveiling the pseudospin-0 character of quasiparticles emerging in the Varma phase.
In this regime, we calculate the Landau levels (LLs) in the presence of a constant magnetic field and emphasize the main differences with the LLs that appear in quasi-relativistic systems on the Lieb lattice \cite{Goldman, Goldman2}. We then determine the emergent Chern-Simons theory \cite{Dunne} in the bulk induced by a weak external electromagnetic field, and derive the Hall conductivity. Surprisingly, an atypical quantum Hall effect is found for the DKP quasiparticles, which emerges from the relativistic nature of the low-energy theory. 

To our knowledge, our work provides the first derivation of a Chern-Simons theory from a microscopic model that is described by a relativistic non-Dirac Hamiltonian. This opens the way to a full quantum-field-theory characterization of topological phases on the Lieb lattice. We propose an implementation of this model in an artificial electronic system, where LL spectroscopy can be employed to reveal the effective electric charge of DKP quasiparticles.

{\bf DKP theory and LLs on the Lieb lattice:--}
We start by considering a tight-binding model on the Lieb lattice with three different species of fermions, as proposed in Ref. \cite{Moore-Varma}. The fermions are represented by the operators $a$, $b$ and $c$ in Fig.~\ref{Fig1}, where $a$ and $c$ represent $p$ orbitals, while $b$ is a $d$ orbital. The momentum-space Hamiltonian of the model is divided in a free and an interacting part, i.e. $H=H_0 + H_{{\rm int}}$, with
\begin{align} \label{tight-binding}
	H_{0} =2 i K\, b^{\dagger}_{\textbf{k}}(s_{x}a_{\textbf{k}}+s_{y}c_{\textbf{k}})-t s_{x}s_{y}a^{\dagger}_{\textbf{k}}c_{\textbf{k}} +\mathrm{h.c.}.
\end{align} 
Here, $K$ and $t$ are real, $s_{x}=\sin(k_{x}/2)$, $s_{y}=\sin(k_{y}/2)$ and
\begin{align} \label{tight-binding-int}
	H_{{\rm int}} = V \sum \,n_{a}n_{c}+\mathrm{h.c.},
\end{align} 
which represents the interaction between $p$ orbitals in the $a$ and $c$ sites, with $n$ denoting the number operator and $V$ the interaction strength. 
By employing a mean-field approximation for the interaction term in Eq.~(\ref{tight-binding-int}) in the particle-hole channel \cite{Moore-Varma}, a new complex hopping term is induced between NNN,
\begin{align} \label{tight-binding-int2}
	H^{{\rm MF}}_{{\rm int}} = im \cos(k_{x}/2)\cos(k_{y}/2)a_{\textbf{k}}^{\dagger}c_{\textbf{k}}+\mathrm{h.c.},
\end{align} 
where $M(k_{x},k_{y})=im \cos(k_{x}/2)\cos(k_{y}/2)= V \langle c^{\dagger}a\rangle$ is the order parameter that behaves like a mass term. For $m=0$, the system exhibits a single Dirac-like cone at the $\Gamma$-point in the first Brillouin zone and a zero-energy flat band.
For $m\neq 0$, the time-reversal symmetry is broken and the system displays an anomalous quantum Hall phase characterized by a non-zero Chern number for the lower band \cite{Moore-Varma}. 

For $t$ smaller than $K$, in the low-energy limit, the corresponding effective Hamiltonian $H_{{\rm DKP}}$ is fully relativistic and reads
\begin{eqnarray}\label{Duffin-Kemmer}
	H_{{\rm DKP}}=K\left[\beta^{1},\beta^{0}\right]\,k_{x}+K\left[\beta^{2},\beta^{0}\right]\,k_{y}+m\beta^{0},
\end{eqnarray}
where the $3\times 3$ matrices $\beta^{\mu}$ with $\mu=0,1,2$ are given by
$$\beta^{0}= \left( \begin{array}{ccc}
0 & 0 & 0 \\
0 &  0 & i \\
0 &  -i & 0 \end{array} \right), \  
\beta^{1}= \left( \begin{array}{ccc}
0 & 0 & -1 \\
0 &  0 & 0 \\
1 &  0 & 0 \end{array} \right),$$ 
$$\beta^{2}= \left( \begin{array}{ccc}
0 & 1 & 0 \\
-1 &  0 & 0 \\
0 &  0 & 0 \end{array} \right).$$
These matrices satisfy the relation
\begin{eqnarray}
	\beta^{\mu}\beta^{\nu}\beta^{\sigma}+\beta^{\sigma}\beta^{\nu}\beta^{\mu}=
	\beta^{\mu}\eta^{\nu\sigma}+\beta^{\sigma}\eta^{\nu\mu},\label{anticommutation}
\end{eqnarray}
where $\eta^{\mu\nu}$ is the Minkowski metric, such that $\mathrm{diag}\,\, \eta^{\mu\nu}=(1,-1,-1)$. The above conditions identify the DKP algebra \cite{Nieto}, which is the core of the DKP theory and describes relativistic pseudospin-0 quasiparticles in two spatial dimensions.

By implementing a Legendre transformation on Eq.~(\ref{Duffin-Kemmer}), the DKP action may be written in terms of a first-order Lagrangian \cite{Nieto}, i.e.
\begin{eqnarray}
S_{{\rm DKP}}[\bar{\psi},\psi]=\int d^{3}x\, \bar{\psi}(i\hbar \beta^{\mu}\partial_{\mu}-m)\psi, \label{action}
\end{eqnarray}
where $d^{3}x=dt d^{2}x$, $\psi=(a,b,c)^{T}$, the adjoint spinor $ \bar{\psi}=\xi^{0}\psi$, with $\xi^{0}=2 (\beta^{0})^{2}-\eta^{00}$, and $K$ has been fixed to unit for simplicity.
Similarly to other relativistic field theories, the spinor field $\psi$ satisfies the Klein-Gordon equation at semiclassical level, i.e. $(\Box+m^{2})\psi=0$.

Now, we proceed by investigating the effect of an external perpendicular magnetic field on the Lieb lattice.
As for the Dirac theory, we introduce an electromagnetic field in the DKP theory by minimally coupling the gauge field $A_{\mu}$ and vector DKP current $j^{\mu}=\bar{\psi}\beta^{\mu}\psi$, i.e. $\partial_{\mu}\rightarrow D_{\mu}=\partial_{\mu}-i (q/\hbar) A_{\mu}$, where $q$ is the electric charge.
We choose the Landau gauge, and determine the LLs in our 2+1-dimensional system, in analogy to Ref.~\cite{Abreu}, where the 3+1-dimensional case is treated. We then find 
\begin{eqnarray}\label{LL}
E_{{\rm LL}} (B)=\pm\sqrt{m^{2}+(2n+1)\hbar \,q B},
\end{eqnarray}
with $n=0,1,2...$. For $m=0$, the LLs coincide with those ones derived in Ref.~\cite{Goldman}, when the (non-dispersing) flat band appears at zero energy. However, for $m\neq 0$, the LLs found here for a fully relativistic theory are different from those of the gapped phase on the Lieb lattice calculated in Ref.~\cite{Goldman} for a quasi-relativistic theory. 

Importantly, Eq.~(\ref{LL}) is equivalent to the LLs of a two-dimensional \textit{scalar} field coupled to an external magnetic field \cite{Schrader}. This unveils the bosonic nature of the DKP field.
\begin{figure}[htp]
	\begin{center}
		\includegraphics[scale=0.45]{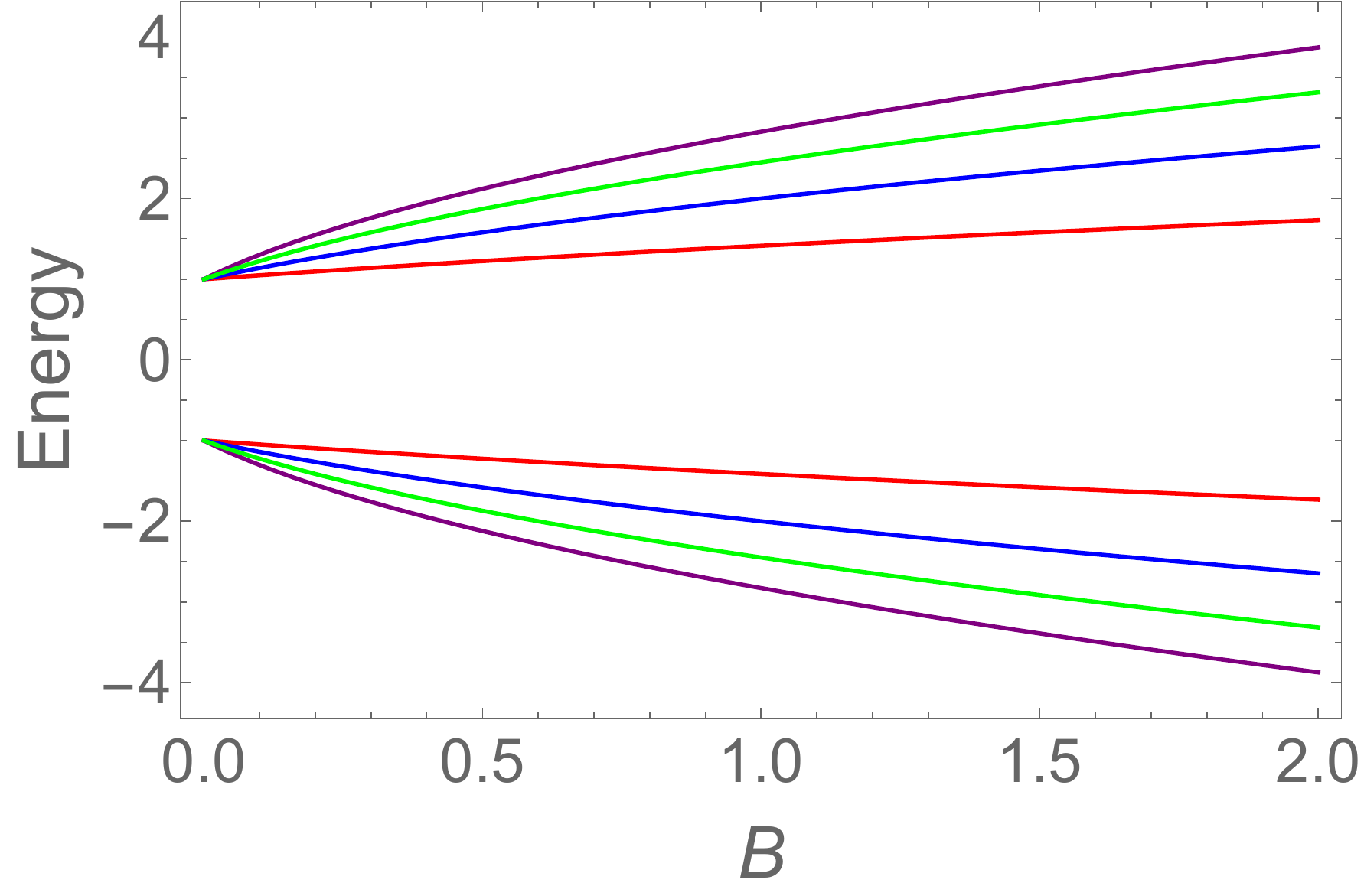}
	\end{center}
	\caption{\label{Fig2} Plot of LLs with respect to the magnetic field B for $n=0,1,2,3$ at fixed mass $m$ and charge $q$.}
\end{figure}

{\bf Chern-Simons theory and Hall conductivity:--}

%State more explictly the difference between this and the previous section.
So far, in the Landau gauge, we have shown that the LL spectrum of the DKP quasiparticles is different from the Dirac case. Now, we investigate, in the weak-field approximation \cite{Fradkin}, the topological response of the system to an external electromagnetic field $A_{\mu} = A_{\mu}(\textbf{x},t)$, which contains both the scalar and the vector potentials. In this way, the gauge field acts like a flux in the Haldane model \cite{Haldane} and yields an anomalous Hall response. Hence, the action (\ref{action}) becomes
\begin{eqnarray}
S_{{\rm DKP}}[\bar{\psi},\psi, A_{\mu}]=\int d^{3}x\, \bar{\psi}(i\hbar\beta^{\mu}\partial_{\mu}+q\beta^{\mu}A_{\mu}-m)\psi.\quad \label{action2}
\end{eqnarray}

The topological behavior is encoded in the effective topological field theory (ETFT) $S_{\rm eff}^{\rm T}[A_{\mu}]$ that can be derived from the above action by integrating out the spinor fields in the corresponding partition function. We have that
\begin{eqnarray}
e^{\frac{i}{\hbar}S_{\rm eff}[A_{\mu}]}=\int D\bar{\psi} D \psi\,\, e^{\frac{i}{\hbar} S_{{\rm DKP}}[\bar{\psi},\psi,A_{\mu}]},
\end{eqnarray}
where the effective action $S_{\rm eff}[A_{\mu}]$ splits into a sum of a topological (T) and a non-topological (NT) contribution,
\begin{eqnarray}
S_{\rm eff}[A_{\mu}]&=&-i \hbar \log \det (i\hbar\beta^{\mu}\partial_{\mu}+q\beta^{\mu}A_{\mu}-m) \nonumber \\
&=& S_{\rm eff}^{\rm T}+S_{\rm eff}^{\rm NT}.  \label{ETFT}
\end{eqnarray}
Like in the two-dimensional massive Dirac theory \cite{Dunne, Redlich,Niemi-Semenoff,Ishikawa}, the ETFT in our case is determined by the calculation of the photon one-loop self-energy diagram and is given by an Abelian Chern-Simons theory
\begin{eqnarray}
S_{\rm eff}^{\rm T}[A_{\mu}]=\frac{1}{2}\int \frac{d^{3}p}{(2\pi)^{3}} A_{\mu}(-p)\Pi^{\mu\nu}(p)A_{\nu}(p),
\end{eqnarray}
where the polarization tensor $\Pi^{\mu\nu}$ is obtained via
\begin{eqnarray}
i\Pi^{\mu\nu}= -\frac{q^{2}}{\hbar}\int \frac{d^{3}k}{(2\pi)^{3}}{\rm Tr} \PR{\beta^{\mu}G_{\psi}(k) \beta^{\nu}G_{\psi}(k-p)}, \label{PO}
\end{eqnarray}
and $G_{\psi}$ is the DKP quasiparticle propagator,
\begin{eqnarray}
G_{\psi}(k) = i \frac{\beta^{\gamma}k_{\gamma}+m}{k^{2}-m^{2}}. \label{propagator}
\end{eqnarray}
Next, we substitute the expression for the propagator above into Eq.~(\ref{PO}), and perform the calculations focusing on the antisymmetric (AS) part of the tensor $\Pi^{\mu\nu}$, which leads to the Chern-Simons term 
\begin{eqnarray} 
i\Pi^{\mu\nu}_{{\rm AS}}=  \frac{q^{2}}{8\pi\hbar}\epsilon^{\mu\nu\alpha}p_{\alpha}{\rm sgn}(m).\label{PiAS}
\end{eqnarray}
Notice that the Levi-Civita tensor $\epsilon^{\mu\nu\alpha}$ arises due to the trace properties of DKP matrices $\beta^{\mu}$, i.e. 
${\rm Tr} [\beta^{\mu}\beta^{\nu}\beta^{\alpha}]=i \epsilon^{\mu\nu\alpha}$. With the result (\ref{PiAS}), by using Kubo's formula, we obtain the Hall conductivity (for $i,j=1,2$) 
\begin{eqnarray}
\sigma^{ij} = \lim_{p_0 \rightarrow 0} \frac{i\Pi^{ij}_{{\rm AS}}}{p_{0}}=  {\rm sgn}(m)\frac{q^{2}}{4h}. \label{Kubo}
\end{eqnarray}

As it stands, Eq.~(\ref{Kubo}) predicts a \textit{quarter-integer} quantum Hall effect for charge $q=e$ fermions. This result is astonishing because we started from the single-particle Hamiltonian (\ref{Duffin-Kemmer}), and as such, one should expect to obtain an integer QHE, even because the Chern number for this model is integer \cite{Palumbo, Moore-Varma}. We explain below the reasons behind this unexpected finding.

%*******************
The knowledge and understanding of the quantum Hall effect in Dirac materials relies on the behavior of Dirac quasiparticles, which are described by the relativistic spin-1/2 Dirac theory. Within this category, we can distinguish three main examples in two spatial dimensions: the Haldane model \cite{Haldane}, the Bernevig-Hughes-Zhang (BHZ) model \cite{BHZ}, and the gapped boundary of three-dimensional topological insulators \cite{Qi-Hughes-Zhang}.

In the first case, there appear two Dirac cones in the bulk due to the Nielsen-Ninomiya theorem (namely, the fermion doubling problem, see Ref.~\cite{Nielsen-Ninomiya}). Each massive Dirac fermion in the bulk with a constant Dirac mass contributes $1/2$ to the total Chern number, such that the bulk Chern number is always an integer. Indeed, the Haldane model supports an anomalous integer quantum Hall effect. 

In the BHZ model, there appears a single Dirac cone in the bulk, and in the gapped phase, the Dirac mass is not constant but momentum-dependent. This is a sort of regularization, where one can avoid the Nielsen-Ninomiya theorem for the Dirac fermion, which induces directly an integer value for the topological invariant in the bulk of the system.

In the case of three-dimensional topological insulators, each gapped surface state supports a two-dimensional massive Dirac fermion leading to a half-integer Chern number, and a half-integer quantum Hall effect \cite{Qi-Hughes-Zhang}. Here, the Nielsen-Ninomiya theorem is avoided because the two-dimensional system lives on the boundary of a higher-dimensional bulk. The Chern number per surface is 1/2 and the Hall conductivity is given in units of $e^{2}/2h$ even if there is no topological ground state degeneracy on the gapped boundary. Moreover, the corresponding effective field theory is given by an Abelian Chern-Simons theory with half-integer level.

The result of Eq.~(\ref{Kubo}) challenges this understanding in two ways: First, the Chern number in the lower band of the DKP model is $\pm 1$ as shown in Refs.~\cite{Palumbo,Moore-Varma}. This implies that the non-integer value of the Hall conductivity in Eq.~(\ref{Kubo}) cannot come from a non-integer value of the Chern number, like in the case of a single massive Dirac fermion mentioned above. On the other hand, the Nilsen-Ninomiya theorem is avoided in our model because of the presence of a flat band \cite{Dagotto}. In other words, there is a single valley and at the same time the DKP mass is also constant in the low-energy limit. Second, the trace of the three $\beta$ matrices is half of the value of the trace of three Pauli matrices.
This trace is related to the anti-symmetric part of the polarization tensor that gives rise to the Chern-Simons action.
These are the two formal reasons behind the fractional coefficient in Eq.~(\ref{Kubo}). 

This original result could nevertheless be reconciled with the integer-valued Hall conductivity if the charge $q=2e$. This picture is consistent with the bosonic nature of our low-energy DKP quasiparticles, which behave effectively like bosons, as revealed by analyzing the properties of their wavefunctions \cite{Bennett}.
Hidden charge $2e$ bosons have been already proposed to appear in the pseudogap regime of high-Tc cuprates, where they emerge from a low-energy theory of a \textit{doped} Mott insulator \cite{Phillips,Phillips2}. 
However, there is no interaction in our Hamiltonian (\ref{Duffin-Kemmer}) that could lead to pairing within a conventional picture of superconductivity.

%This is, in our opinion, the main interest of our result: it shows that this "common sense" does not apply to more generic relativistic models, and we can immediately understand why this occurs, in mathematical terms, by looking at the trace of the beta matrices, which is different from the trace of Dirac matrices, and at the lack of fermion doubling in the three-band model. 

%*******************

%The presence of the additional factor $1/4$, in comparison to the Dirac case, can be understood as follows: Firstly, the trace of the three $\beta$ matrices is half of the value of the trace of three Pauli matrices; secondly, the valley (i.e. the fermion doubling), which appears in graphene and related honeycomb materials, is absent due to the presence of the flat band. However, it is interesting to notice that for $q=2e$, Eq.~(\ref{Kubo}) yields an integer quantum Hall effect. 

{\bf Edge modes:--} The existence of topologically protected chiral edge modes on the Lieb lattice has been shown in Ref.~\cite{Palumbo} by employing the entanglement spectrum. Here, we provide an analytic derivation of the Dirac edge modes from the DKP Hamiltonian by following a domain-wall argument. In particular, we employ the Jackiw-Rebbi approach \cite{Jackiw}, as already done in Dirac systems.

The boundary of the system corresponds to a nodal interface at $x=0$, where the mass $m(x)$ runs from negative to positive (or vice-versa).
Let us consider the DKP Hamiltonian in Eq.~(\ref{Duffin-Kemmer}) with a spatially varying mass term, and for simplicity we set $K=1$ and $\hbar=1$. Importantly, we assume that $m(x)_{x\rightarrow \pm \infty}=\pm m_{0}$. Notice that due to the periodicity in the $y$-direction, $k_{y}$ is still a good quantum number. Therefore, we can replace $\partial_{y}\rightarrow ik_{y}$ in the Hamiltonian, such that the eigenvalue problem $H_{{\rm DKP}}\Psi=E\Psi$ can be solved by solutions $\Psi(x,y)=(u(x),v(x),w(x))^{\intercal}e^{ik_{y}y}$
with energy $E(k_{y})=k_{y}$.
We then obtain the following first-order differential equations
\begin{eqnarray}
-k_{y}u+\partial_{x}v+ik_{y}\, w&=&0, \nonumber \\
-\partial_{x}u-k_{y}v+im\, w&=&0, \nonumber \\
-ik_{y}\, u-i m\,v-k_{y}w&=&0.\label{domain-wall}
\end{eqnarray}
The last equality leads to $u=i(k_{y}w+i m\,v)/k_{y}$, which may be substituted into the first equation to yield a first-order 
differential equation for $v(x)$
\begin{eqnarray}
\partial_{x}v(x)=-m(x) v(x). \label{Ux}
\end{eqnarray}
The solution of Eq.~(\ref{Ux}) reads $v(x)= c_{1}\, e^{-\int dx\,m(x)}$, with $c_{1}$ a suitable real constant. In the simplest case, in which the domain wall is described by a Heaviside step function, i.e. $m(x)=m_{0}[2\theta(x)-1]$, we obtain 
\begin{eqnarray}
v(x)= c_{1} e^{-m_{0}[2\theta(x)-1] x } = c_{1} e^{-m_{0}|x|}. \label{vsolution}
\end{eqnarray}
which represents a localized edge state at $x=0$ when $m(x)$ goes from positive to negative, i.e. $m_{0}$ is positive and Eq.~(\ref{vsolution}) is a normalizable wave function.
From Eqs.~(\ref{domain-wall}), we also obtain 
\begin{eqnarray}
\partial_{x}u(x)-m(x)u(x) =\left(\frac{m_{0}^{2}-k_{y}^{2}}{k_{y}}\right)v(x), \label{Vx}
\end{eqnarray}
where the equality holds because $\theta(x)^{2}=\theta(x)$. Thus, the solution of Eq.~(\ref{Vx}) is given by
\begin{eqnarray}
u(x)=\left(\frac{m_{0}^{2}-k_{y}^{2}}{k_{y}}\right)e^{\int dx\, m(x)}\int dx\, e^{-2\int dx\, m(x)}, \label{usolution}
\end{eqnarray}
which does not describe any localized edge mode at $x=0$. It can be easily shown that $u(x)_{x\rightarrow 0}=0$ and $u(x)_{x\rightarrow \pm \infty}=\pm\infty$. At the same time, $w(x)$ does not describe any edge mode either. 
However, when $m(x)$ goes from negative to positive, $m_{0}$ is negative and Eq.~(\ref{vsolution}) does not correspond any physical solution because it yields a non-renormalizable solution. In this case, $c_{1}$ is zero, such that $v(x)=0$. By plugging this solution into Eqs.~(\ref{domain-wall}), we get different solutions for $u(x)$ and $w(x)$. In this case, we find that $u(x)= c_{2}\, e^{\int dx\,m(x)}$ (with $c_{2}$ a suitable real constant), which identifies a localized edge mode because $m_{0}<0$. On the other hand, we get also $w(x)=-i u(x)$, which does not correspond to any localized mode in the real plane. 

Thus, in both cases, i.e. $m_{0}>0$ and $m_{0}<0$, we can show that there is always a single propagating edge mode along the domain wall at $x=0$. This single Dirac edge mode propagating on the boundary of the system may be represented by a scalar field in 1+1-dimensions. Therefore, in this system, the DKP theory exists only in 2+1-dimensions, and is absent at the edge.   

\begin{figure}[t!]
	\begin{center}
		\includegraphics[scale=0.15]{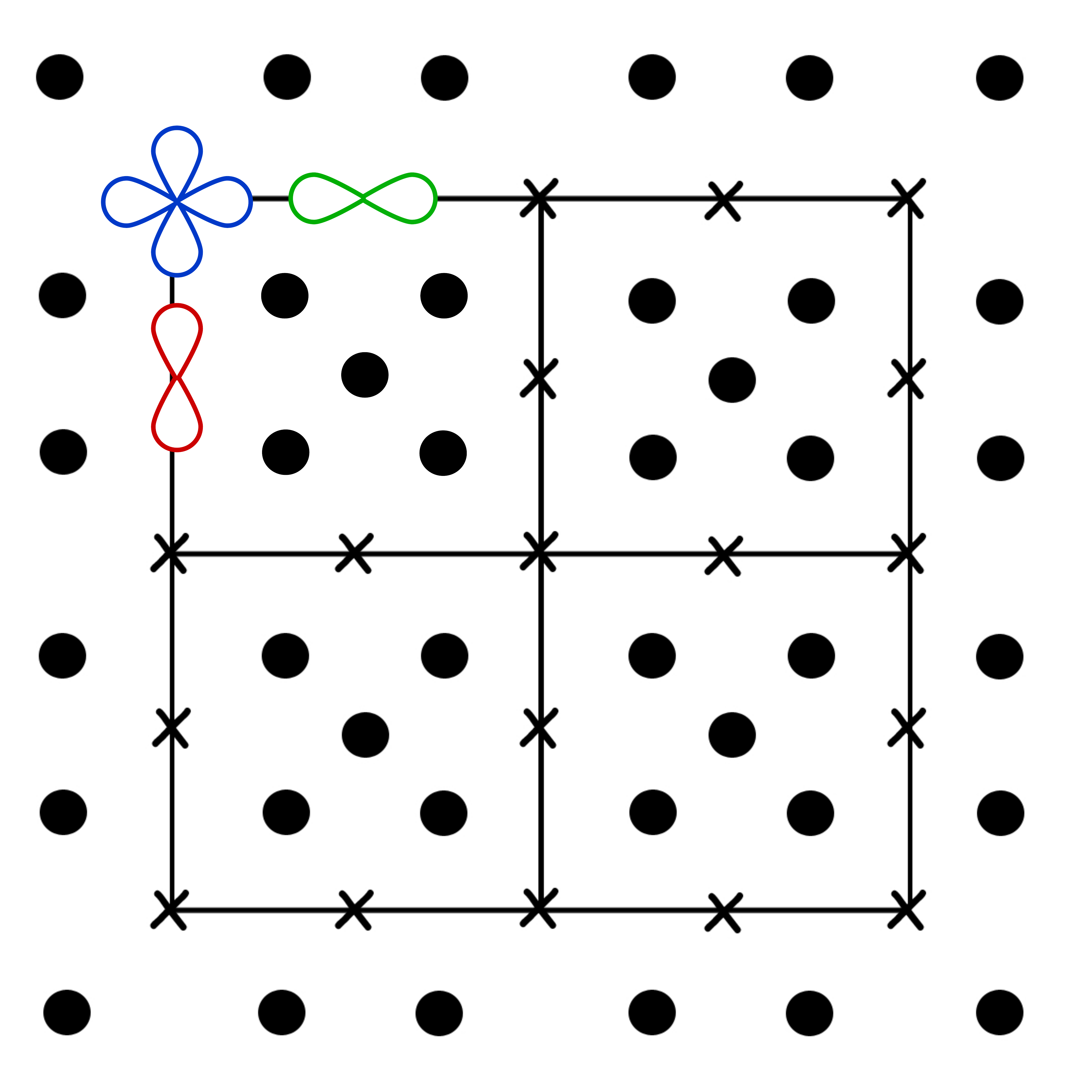}
	\end{center}
	\caption{\label{Fig3} Patterning of an electronic Lieb lattice using CO adsorbates.}
\end{figure}

{\bf Experimental realization:} Here, we propose an electronic quantum simulation of the Cu-O model analyzed in this paper. By following Ref.~\cite{Morais-Smith2}, we consider an electronic Lieb lattice built by confining the electrons on the surface state of Cu(111) by an array of CO molecules positioned with a scanning tunneling microscope (STM), as shown in Fig.~\ref{Fig3}. The CO molecules are represented by black circles, which act as a repulsive potential and confine the electrons to the Lieb-lattice geometry. To simulate, respectively, the $d$ and $p$-orbitals at the corner and side sites, we propose to use an anisotropic configuration, in which the four COs are further away from the corner, and closer to the side sites, as depicted in Fig.~\ref{Fig3}. This leads to weaker confinement at the corner, and hence to higher orbital wave functions. 

By investigating the behavior of the system at low temperatures, under ultra-clean conditions, we expect to reach the interacting regime, where the mean-field Varma phase with loop currents in each plaquette should emerge. Similar systems have been emulated with ultra cold bosons in optical lattices \cite{Hemmerich1,Hemmerich2}. Finally, an external magnetic field will be added perpendicularly to the Lieb lattice, to induce LLs and the quantum Hall state in the system.

In this electronic quantum simulator, the value of the quasiparticle charge may be determined by spectroscopic measurements of the LLs. This procedure may provide a direct observation of the charge $q$ in the Varma phase, and resolve the puzzling result obtained for the Hall conductivity.

{\bf Conclusions:--}
In this letter, we investigate the electromagnetic response of pseudospin-0 DKP quasiparticles on the Lieb lattice. First, we derive the LLs induced by a constant magnetic field orthogonal to the system. Then, we analyze the topological response of the model in presence of a U(1) gauge potential in terms of an Abelian Chern-Simons theory, from which we obtained an atypical Hall conductivity. The experimental realization of this setup will reveal the real effective value of the electric charge of DKP quasiparticles. In any case, the relativistic description of non-Dirac systems holds promises to reveal a much richer physics than in conventional Dirac materials.

{\bf Acknowledgments:--}
 This work was supported by CNPq (Brazil) through the Brazilian government project Science Without Borders. We are grateful to V. Gritsev, J. A. Hutasoit, T. H. Hansson, A. Hemmerich, J. van Wezel, M. Di Liberto, E. C. Marino and I. Swart for inspiring discussions. \\


\begin{thebibliography}{100}


\bibitem{Novoselov}
K.S. Novoselov, A.K. Geim, S.V. Morozov, D. Jiang, M.I. Katsnelson, I.V. Grigorieva, S.V. Dubonos and A.A. Firsov,
Nature \textbf{438}, 197, (2005).

\bibitem{Kane}
M. Z. Hasan and C. L. Kane,
Rev. Mod. Phys. \textbf{82}, 3045 (2010).

\bibitem{Zhang}
X.-L. Qi and S.-C. Zhang
Rev. Mod. Phys. \textbf{83}, 1057 (2011).

\bibitem{Fradkin} 
E. Fradkin, Field Theories of Condensed Matter Physics, (Cambridge University Press, 2013).

\bibitem{MacDonald}
D. A. Pesin and A. H. MacDonald,
Nature Materials \textbf{11}, 409 (2012).

\bibitem{Pachos} 
J. K. Pachos, Introduction to Topological Quantum Computation, (Cambridge University Press, 2012).

\bibitem{Ramond} 
P. Ramond, Field Theory: A Modern Primer, (Hachette, 1997).

\bibitem{Rappe}
S. M. Young, S. Zaheer, J. C. Y. Teo, C. L. Kane, E. J. Mele, and A. M. Rappe,
Phys. Rev. Lett. \textbf{108}, 140405 (2012).

\bibitem{Turner}
X. Wan, A. M. Turner, A. Vishwanath, and S. Y. Savrasov,
Phys. Rev. B \textbf{83}, 205101 (2011).

\bibitem{Nakahara} 
M. Nakahara, Geometry, Topology and Physics (CRC Press, 2003).

\bibitem{Altland} 
A. Altland and M. R. Zirnbauer,
Phys. Rev. B \textbf{55}, 1142 (1997).

\bibitem{Ryu2} 
A. P. Schnyder, S. Ryu, A. Furusaki, and A. W. W. Ludwig,
Phys. Rev. B \textbf{78}, 195125 (2008).

\bibitem{Kitaev} 
A. Kitaev, AIP Conf. Proc. \textbf{1134}, 22 (2009).

\bibitem{Ryu} 
C.-K. Chiu, J. C. Y. Teo, A. P. Schnyder, and S. Ryu,
Rev. Mod. Phys. \textbf{88}, 035005 (2016).


\bibitem{Bercioux} 
D. Bercioux, D. F. Urban, H. Grabert, and W. Häusler
Phys. Rev. A \textbf{80}, 063603 (2009).

\bibitem{Shen} 
R. Shen, L. B. Shao, B. Wang, and D. Y. Xing,
Phys. Rev. B \textbf{81}, 041410(R) (2010).

\bibitem{Green} 
D. Green, L. Santos, and C. Chamon,
Phys. Rev. B \textbf{82}, 075104 (2010).

\bibitem{Franz} 
C. Weeks and M. Franz,
Phys. Rev. B \textbf{82}, 085310 (2010).

\bibitem{Moessner} 
B. Dora, J. Kailasvuori, and R. Moessner,
Phys. Rev. B \textbf{84}, 195422 (2011).

\bibitem{Goldman} 
N. Goldman, D. F. Urban, and D. Bercioux,
Phys. Rev. A \textbf{83}, 063601 (2011).

\bibitem{Goldman2} 
Z. Lan, N. Goldman, A. Bermudez, W. Lu, and P. Ohberg,
Phys. Rev. B \textbf{84}, 165115 (2011).

\bibitem{Morais-Smith} 
W. Beugeling, J. C. Everts, and C. Morais Smith,
Phys. Rev. B \textbf{86}, 195129 (2012).

\bibitem{Orlita} 
M. Orlita, D. M. Basko, M. S. Zholudev, F. Teppe, W. Knap, V. I. Gavrilenko, N. N. Mikhailov, S. A. Dvoretskii, P. Neugebauer, C. Faugeras, A.-L. Barra, G. Martinez and M. Potemski,
Nat. Phys. \textbf{10}, 233 (2014).

\bibitem{Bernevig} 
B. Bradlyn, J. Cano, Z. Wang, M. G. Vergniory, C. Felser, R. J. Cava, and B. A. Bernevig, Science \textbf{353}, (2016).

\bibitem{Peotta} 
A. Julku, S. Peotta, T. I. Vanhala, D.-H. Kim, and P. Torma,
Phys. Rev. Lett. \textbf{117}, 045303 (2016).

\bibitem{Soluyanov} 
Z. Zhu, G. W. Winkler, Q. Wu, J. Li, and A. A. Soluyanov,
Phys. Rev. X \textbf{6}, 031003 (2016).

\bibitem{Assaad} 
M. Bercx, J. S. Hofmann, F. F. Assaad, and T. C. Lang,
Phys. Rev. B \textbf{95}, 035108 (2017).

\bibitem{Stern} 
I. C. Fulga and A. Stern,
Phys. Rev. B \textbf{95}, 241116(R) (2017).

\bibitem{pseudo1} 
This pseudospin-1 characterizes an internal degree of freedom of the wavefunctions -- as for layers, polarizations, sublattices etc.




\bibitem{Morais-Smith2}
M. R. Slot,	T. S. Gardenier, P. H. Jacobse,	G. C. P. van Miert,	S. N. Kempkes, S. J. M. Zevenhuizen,	C. Morais Smith, D. Vanmaekelbergh	and I. Swart,
Nat. Phys. \textbf{13}, 672 (2017).


\bibitem{Ojanen}
R. Drost, T. Ojanen, A. Harju and P. Liljeroth,
Nat. Phys. \textbf{13}, 668 (2017).

\bibitem{Vicencio}
D. Guzman-Silva, C. Mejia-Cortes, M. A. Bandres, M. C.
Rechtsman, S. Weimann, S. Nolte, M. Segev, A. Szameit
and R. A. Vicencio, New J. Phys. \textbf{16}, 063061 (2014).

\bibitem{Goldman3}
S. Mukherjee, A. Spracklen, D. Choudhury, N. Goldman, P. Ohberg, E. Andersson, and R. R. Thomson,
Phys. Rev. Lett. \textbf{114}, 245504 (2015).

\bibitem{Diebel}
F. Diebel, D. Leykam, S. Kroesen, C. Denz, and A. S. Desyatnikov,
Phys. Rev. Lett. \textbf{116}, 183902 (2016).

\bibitem{Palumbo} 
G. Palumbo and K. Meichanetzidis,
Phys. Rev. B \textbf{92}, 235106 (2015).

\bibitem{Nieto}
R. A. Krajcik and M. M. Nieto, Phys. Rev. D \textbf{10}, 4049
(1974).

\bibitem{Varma}
C. M. Varma, Phys. Rev. B \textbf{55}, 14554
(1997).

\bibitem{Varma1}
C. M. Varma, Phys. Rev. B \textbf{73}, 155113
(2006).


\bibitem{Moore-Varma}
Y. He, J. Moore and C. M. Varma, Phys. Rev. B \textbf{85}, 155106
(2012).

\bibitem{Dunne}
G. V. Dunne, Les Houches Lectures 1998, arXiv:hep-th/9902115.


\bibitem{Abreu}
L. M. Abreu, E. S. Santos and J. D. M. Vianna,
J. Phys. A: Math. Theor. \textbf{43}, 495402 (2010).

\bibitem{Schrader}
R. Schrader, Fortsch. Phys. \textbf{20}, 701 (1972).


\bibitem{Haldane} 
F. D. M. Haldane, Phys. Rev. Lett. \textbf{61}, 2015 (1988).

\bibitem{Redlich}
A. N. Redlich, Phys. Rev. D \textbf{29}, 2366 (1984).

\bibitem{Niemi-Semenoff}
A. Niemi and G. Semenoff, Phys. Rev. Lett. \textbf{51}, 2088 (1983).  

\bibitem{Ishikawa}
K. Ishikawa, Phys. Rev. Lett. \textbf{53}, 1615 (1984).

\bibitem{BHZ}
B. A. Bernevig, T. L. Hughes and S.-C. Zhang, Science \textbf{314}, 1757 (2006).

\bibitem{Qi-Hughes-Zhang} 
X.-L. Qi, T. L. Hughes and S.-C. Zhang, Phys. Rev. B \textbf{78}, 195424 (2008).

\bibitem{Nielsen-Ninomiya} 
H. B. Nielsen and M. Ninomiya, Phys. Lett. B \textbf{105}, 219 (1981).

\bibitem{Dagotto} 
E. Dagotto, E. Fradkin and A. Moreo, Phys. Lett. B \textbf{172}, 383 (1986).

\bibitem{Bennett}
A. F. Bennett, Found. Phys. \textbf{46}, 1090 (2016).

\bibitem{Phillips}
S. Chakraborty and P. Phillips, Phys. Rev. B \textbf{80}, 132505 (2009).

\bibitem{Phillips2}
R. G. Leigh, P. Phillips and T.-P. Choy, Phys. Rev. Lett. \textbf{99}, 046404 (2007).

\bibitem{Jackiw}
R. Jackiw and C. Rebbi,
Phys. Rev. D \textbf{13}, 3398 (1976).

\bibitem{Hemmerich1}
G. Wirth, M. Olschlager	and A. Hemmerich,
Nat. Phys. \textbf{7}, 147 (2011).

\bibitem{Hemmerich2}
M. Olschlager, T. Kock, G. Wirth, A. Ewerbeck, C. Morais Smith and A. Hemmerich,
New J. Phys. \textbf{15}, 083041 (2013).


\end{thebibliography}
\end{document}